\documentclass{aa}
\usepackage{graphicx}

\begin{document}

\title{Bias and consistency in
time delay estimation methods: case 
of the double quasar HE 1104-1805}

\author{J. Pelt \inst{1,2} \and
        S. Refsdal \inst{2,3} \and
        R. Stabell \inst{2}}

\offprints{J.Pelt, e-mail: {pelt@aai.ee}}

\institute {Tartu Observatory, T\~oravere, 61602 Estonia
           \and 
           Institute of Theoretical Astrophysics, University of Oslo,
           P.O. Box 1029, Blindern, N-0315 Oslo, Norway
           \and
           Hamburger Sternwarte, Gojenbergsweg 112, D-21029, 
           Hamburg-Bergedorf, Germany
           }

\date{Received; accepted}

\authorrunning{J.\ Pelt}
\titlerunning{Double quasar HE 1104-1805 time delay}

\abstract{
We present a short re--evaluation of a recently published
time delay estimate for the gravitational lens system HE 1104--1805
with emphasis on important methodological aspects:
bias of the statistics, inconsistency of the methods
and use of the 
purposeful selection of data points
(or so--called ``cleaning'') at the
preprocessing stage.
We show how the inadequate
use of simple analysis methods can lead to
too strong conclusions.
Our analysis shows that there are indications for the time delay
in HE 1104--1805 to be between $-0.9$ and $-0.7$ years, but 
still with a large uncertainty.

}
\maketitle

\section{Introduction}

In a recent paper by Gil-Merino et al.~(\cite{GilMerino02}, 
in the following referred to as GWW)
photometric data for the double quasar HE 1104-1805 was used to get
a new and improved estimate for the time delay of the system.
The authors claim that the previous value obtained 
for the delay ($-0.73$ years, Wisotzki et al.~\cite{Wisotzki98}) was
affected by bias
of the employed dispersion minimization method (Pelt
et al.~\cite{Pelt94,Pelt96}). To
substantiate their claim they introduce a notion of 
{\it consistency} to evaluate
different delay estimation schemes.

When the bias (and methods to control it) of the dispersion minimization
algorithm is well known, its possible 
internal inconsistency is certainly a new aspect
of the method and should be seriously considered. In this short note we
just do that. First we shortly review the problem of the bias introduced
by the method and then turn to the problem of inconsistency. We will show
that this notion as defined in GWW comes 
from a shaky argument and should not be used.
This is true in the context of the dispersion minimization method
as well as in the context of other methods.

In the final part of the paper we will show that the currently available
data for HE 1104-1805 is not sufficient to determine
a time delay with a resonable precision (with a standard
deviation of, say, $\leq 15\%$).
 
\section{Bias in the dispersion minimization method}
In the dispersion minimization method 
(Pelt et al.~\cite{Pelt94,Pelt96})
one evaluates cross sums of the
type:
\begin{displaymath}
D^2(\tau)={\sum_{n=1}^{N-1}\sum_{m=n+1}^{N} W_{n,m}(\tau)(C_n-C_m)^2 \over
\sum_{n=1}^{N-1}\sum_{m=n+1}^{N} W_{n,m} },
\end{displaymath}
where 
$C_n$ and $C_m$ are points of the combined lightcurve 
and 
the weights $W_{n,m}(\tau)$ depend on 
the trial time delay $\tau$, estimated errors and a particular
downweighting scheme $S(t_n-t_m)$ (for details see Pelt et al. \cite{Pelt94, Pelt96}
or GWW). Bias is introduced into the method
by the fact that magnitude differences 
$C_n-C_m=q_n+\epsilon_n-q_m-\epsilon_m$ 
depend not only on
observational errors $\epsilon_n,\epsilon_m$ 
but also on the intrinsic variability of the quasar through $q_n-q_m$.
The basic presumption of the method is that this variability is
relatively low for nearby time points which have non-zero weights in
the above sum. Bias in the method
is somewhat controlled by the choice of the parameter 
$\delta$ in a particular downweighting scheme
\begin{displaymath}
S(t_n-t_m)=
\left\{
\begin{array}{ll}
1-{|t_n-t_m| \over \delta} & \mbox{if  $ |t_n-t_m|\le \delta $} \\
0 & \mbox{if  $ |t_n-t_m|> \delta $}, 
\end{array}
\right.
\end{displaymath}
which gives a higher
weight for nearby pairs and lower weight for pairs whose components
are far away from each other. Pairs with 
$|t_n-t_m|>\delta$ are excluded. 
To use the dispersion method consistently we need to compute dispersion
spectra for a large sample of downweighting parameter $\delta$ values. 
If there are
enough points in the data set to compute
statistically stable dispersions and they are favorably spaced 
(without long regularly spaced gaps) then the best
delay shows up in  a multitude of spectra. The scatter of 
the obtained minima can be 
used to get a rough estimate of the bias involved. We refer 
the reader to
Pelt et al.~(\cite{Pelt96}) for a detailed analysis of the bias in 
the dispersion minimization method as used in the case of 
the double quasar  
QSO 0957+561.

For the analysis below we note that
dispersion minimization can be looked upon as a certain election
scheme. The pool of electors (or voters)
is a set of all data point pairs $(C_n,C_m)$ in
a combined curve ($A$--curve together with the time and 
magnitude shifted $B$--curve). We assume here that the $C_n$ and 
$C_m$ always originate from different original curves (one from $A$ and
the other from $B$). For a fixed trial time delay 
$\tau_{\rm trial}$ and a magnitude shift optimized for this value, 
every pair gives a vote of a certain
strength. The  vote is
zero when  $|t_n-t_m|> \delta$ for a certain pair.
Let us call such electors {\it irrelevant} for a
particular time delay. The strength of the other electors' vote depend on
the magnitude difference $C_n-C_m$ and on the weight applied. 
If, for instance,
there is a pair (elector) whose time difference for a time delay
$\tau_{\rm trial}$ is quite small, and whose squared magnitude 
difference is high, then this elector
gives a very strong vote {\em against} $\tau_{\rm trial}$.
The trial time delay
which gathers the smallest amount of against votes is the winner.
The dispersion spectrum is just a voting protocol for the row
of trial time delays.

\section{Consistency of the time delay estimation methods}
The notion of consistency for
time delay estimation schemes was introduced in GWW. 
It differs from that usually used
in mathematical statistics, so we need to give some details. 
Using the metaphore of elections introduced above, the basic idea
behind their ``consistency'' can be explained as follows.
Let us first compute the dispersion spectrum. Assume that
it attained a global minimum at time delay $\tau_1$. We can look now 
at our voters. For this particular delay a certain subset of voters
is irrelevant (set of zero weight voters). And some 
of the original data points take part only
in zero votes. We remove these irrelevant points  
from the data set and recompute the spectrum (second tour
of the elections). 
If the new spectrum has a strongest
minimum for a significantly different delay (say $\tau_2$) then 
the full estimation scheme is {\em inconsistent} according to GWW (Sec. 3.2).

Let us now return to the first tour of our elections. Some of the
electors voted against delay $\tau_1$ and some others
against delay $\tau_2$. 
On average, the first time delay received a smaller
number of ``against'' votes and therefore it won the race.
Now, if
we throw away the {\em irrelevant} points and corresponding voters for 
the delay $\tau_1$ then we often throw
away some of the {\em against} votes for $\tau_2$. It is not
a miracle then that the second delay can win at the second tour! 
There are not
any internal logical or statistical inconsistencies involved.
The important point here is that the best candidate for the true time delay
is found by comparing dispersions for different $\tau$-s and these
dispersions depend on different subsets of data point pairs.
Even when a particular pair is not numerically accounted for to compute
the dispersion at $\tau_1$, 
it can take part
in forming the dispersion for 
another delay $\tau_2$ and thus can determine
which of the delays will be the best estimate.    

What is the probability that the consistency argument
leads to 
a wrong conclusion?
A comprehensive analysis of this question is out of the scope for this
paper. However, a simple numerical experiment shows 
that we are not dealing
with a rare event. Using the HE 1104--1805 data set we 
constructed an artificial
``source curve'' and generated from it 
1000 new data sets as in normal
bootstrap calculations. The only difference was 
that we did not use 
full amplitude bootstrap errors but scaled them by
an error level parameter $R$. 
In Table~\ref{table1} we present the results
of such experiments with different 
parameters $R$, $\delta$ and $\tau$
which model the actual situation with HE 1104-1805.
The {\it Hits} column counts random runs 
whose best delay differed from
the true delay by no more that $0.1$ years. The {\it Rejected} column counts
correctly estimated delays which were 
nevertheless rejected using the consistency argument
(the correctly computed delay jumped more than 
$0.1$ years after ``cleaning'' was applied). 
As we can see,
even for a statistically more stable case with $\delta = 0.35$ the
false rejection rate is considerable for a range of error levels.
\begin{table}
\begin{flushleft}
\begin{tabular}{rrrrr}
\hline\noalign{\smallskip}
& \multicolumn{2}{c}{$\tau=-0.84$ years} & \multicolumn{2}{c}
{$\tau=-0.71$ years} \\
& \multicolumn{2}{c}{$\delta=0.14$} & \multicolumn{2}{c}{$\delta=0.35$} \\
$R$ & {\rm Hits} & {\rm Rejected} & {\rm Hits} & {\rm Rejected} \\ 
\noalign{\smallskip}
\hline\noalign{\smallskip}
1.0  & 99   & 55   & 317   & 19 \\
0.5  & 168  & 79   & 383   & 7 \\
0.2  & 491  & 164  & 592   & 4 \\
0.1  & 758  & 158  & 840   & 1 \\
0.05 & 940  & 108  & 986   & 0 \\
0.02 & 1000 & 11   & 1000  & 0 \\
0.01 & 1000 & 0    & 1000  & 0 \\
\noalign{\smallskip}
\hline
\end{tabular}
\end{flushleft}
\caption[]{Hit and false rejection rates for the two 
series of model calculations. The model 
``source curve'' was constructed by five point median 
filtering of the combined
HE 1104--1805 A curve and time shifted B curve. 
The level of the redistributed
errors was controlled by parameter $R$. There were 
1000 randomization runs in every experiment. 
{\it Hits}\  counts events when
a random run recovered the true time delay with $0.1$ year precision.
{\it Rejected} counts cases when a correctly recovered time delay was
rejected using the consistency argument.  
} 
\label{table1}
\end{table}

Here we used the dispersion method to analyse the notion of
consistency, but all this applies equally well to the methods based on 
a discrete correlation
function. There we should only speak about ``for'' votes instead of 
``against'' votes.

The comparison of the different time delay 
estimation schemes in  GWW is based on this 
(not applicable) notion of consistency.
From the available 38 measured
values of HE 1104-1805 magnitudes the authors 
throw away up to 9 points (24\%) assuming
that such ``cleaning'' , if it results in consistent
spectra (strongest minimum remains approximately in place),
allows to ``eliminate border effects and gaps''. 

\section{What then about the time delay for HE 1104-1805?}
Before describing our own analysis of the HE 1104-1805 data
set let us note that the authors of the GWW are somewhat
inconsistent in their analysis from the very start. 
\begin{figure}
\resizebox{\hsize}{!}{\includegraphics{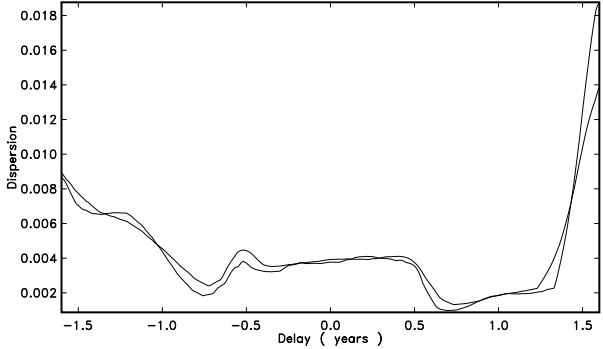}}
\caption[]{
Dispersion spectra with $\delta=0.3$ years (thin line) and
$\delta=0.4$ years (thick line). This is analoguous to Fig.~3 in GWW, only
computed for a symmetric range and with different scaling.}
\label{figr01}
\end{figure}
When the
dispersion spectrum is computed for the full 
symmetric range of the time delays $[-1.6,1.6]$ (years)  
it is seen in Fig.~\ref{figr01} that the strongest minimum occurs 
at a positive delay
around $0.7$ years (A leading B). 
The authors plot only part of the spectra in the asymmetric 
range $[-1.6,0.6]$ years without any specific explanation. From analyses
of the system geometry
by Remy et al.~(\cite{Remy98}) and Wisotzki et al.~(\cite{Wisotzki98}) 
it can, however, be concluded that B should lead A and consequently it is
resonable to seek only negative delays. This is why we restrict
our analysis below to the range of negative delays as well. Nevertheless, we
consider the minimum at $\tau = 0.7$ years as a strong 
warning signal about possible
problems with the time delay estimation for HE 1104-1805.

\begin{figure}
\resizebox{\hsize}{!}{\includegraphics{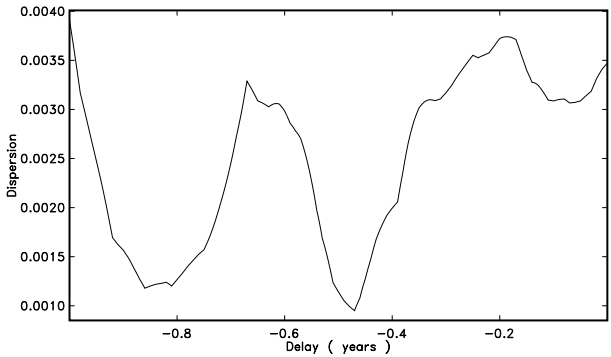}}
\caption[]{
Dispersion spectrum for $\delta=0.14$ years. The left minimum at $-0.86$
years nearly exactly corresponds to the solution obtained in GWW. The middle
mimimum results from the extreme sparseness of the data set for 
delays between $-0.65$ and  $-0.4$ years.  
}
\label{figr02}
\end{figure}

\begin{figure}
\resizebox{\hsize}{!}{\includegraphics{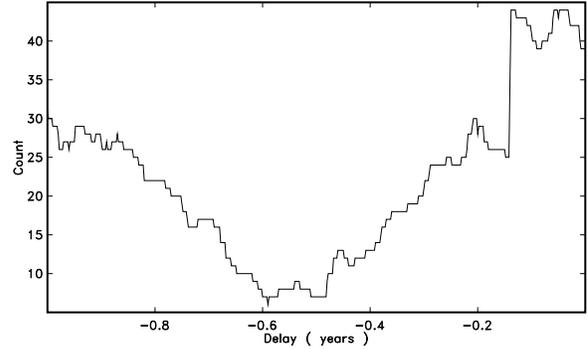}}
\caption[]{
``Window function'' for the dispersion spectrum with $\delta=0.14$ years.
The number of data point pairs used for dispersion estimation (and as well
for correlation function computations) is extremely low in the
region from $-0.65$ to $-0.4$ years.
}
\label{figr03}
\end{figure}

To consistently compare the methods used in GWW with the dispersion
minimization method we first computed a simple spectrum
with linear downweighting parameter $\delta=0.14$ years which corresponds
to the semiwidth $0.07$ years used in GWW. 
As seen from Fig.~\ref{figr02}
there are two strong minima in the spectrum: at 
$-0.86$ and $-0.47$ years. 
The left minimum nearly exactly corresponds to the solution obtained
in GWW. The right (and somewhat stronger) minimum is certainly
a result of sparseness of the data. Just look at Fig.~\ref{figr03} where
a so called ``window function'' for the dispersion spectrum is depicted.
It shows for every particular trial time delay how many data point pairs
have non-zero value (relevant voters!). 
We can easily see that the data is nearly totally blind for the 
time delays between  $-0.65$ and $-0.4$ years. 

It is interesting to recall that
the well known data set of the double quasar QSO 0957+561 which was used
in  Press et al.~(\cite{Press92})
and Pelt et al.~(\cite{Pelt94}), consisted
of 131 time points and its data window minimum that occured around $535$ days
had 43 pairs to compute dispersion from. 
By now it is generally accepted that the correct time delay
in QSO 0957+561 is about 420 days (Kundi\'c et al.~\cite{Kundic97}).
Nevertheless, the 
sophisticated analysis by Press et al.~(\cite{Press92}) did
result in the presumably {\it wrong} value around the minimum of the data
window.
In the case of HE 1104-1805 we have only 19
time points and there are fewer than 10 pairs
to compute the dispersion (or correlation) for a large range of time delays! 

\begin{figure}
\resizebox{\hsize}{!}{\includegraphics{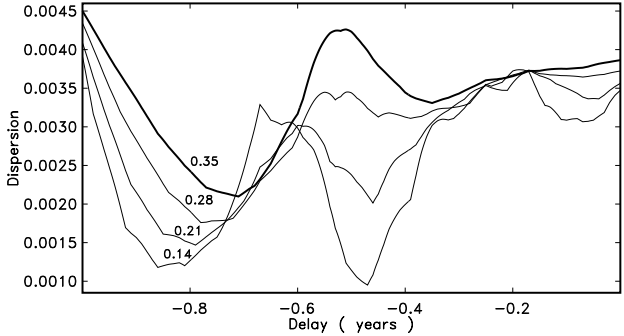}}
\caption[]{
Set of dispersion spectra
for HE 1104--1805 computed with $\delta = 0.14,0.21,0.28,0.35$ years
(increasing line thickness).
It is
seen that when moving towards the longer values of $\delta$ the spectra
become smoother and statistically more stable, the artefact of
data sparseness around $-0.47$ years disappears and the minima around
$-0.8$ years show increasing (probable) 
shift due to the oversmoothing.
}
\label{figr04}
\end{figure}

But probably most revealing is the set of spectra computed with
$\delta = 0.14, 0.21, 0.28 , 0.35$ years and depicted all
together in Fig.~\ref{figr04}.
One can see how the spectra tend to become
smoother and statistically more stable with increasing $\delta$,
how the fluctuation around $-0.47$ years gradually disappears,
and most importantly, how the minimum at $-0.86$ years moves first
to $-0.79$ years, then to $-0.78$ years and finally to $-0.71$
years. This shift can be just
a result of the bias which is introduced due the oversmoothing.
It is also important to note that this consistent picture is obtained
without any ``cleaning'' of the original data set.

\begin{figure}
\resizebox{\hsize}{!}{\includegraphics{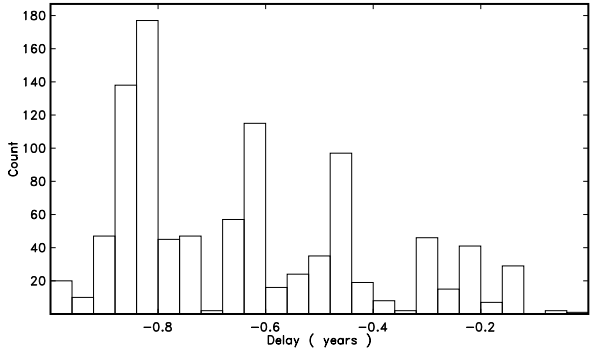}}
\caption[]{
Distribution of the 1000 bootstrap runs 
with $\delta = 0.14$ years and  $\tau = -0.86$ years. 
Because of data sparseness
it was possible to use only a 3 point median filter to get
a rough estimate for a smooth combined curve.
Even for this, certainly conservative scheme, we can see that practically all
possible delays resulted from the trial runs. The main maximum 
of the distribution (mode) occurs near the input delay but it is
not strong enough to give a
support for $\approx -0.85$ year solution.
}
\label{figr05}
\end{figure}

Can we now claim that the time delay 
estimate around $-0.85$ years is reasonably well established
because it shows itself somehow (after ``cleaning'' of data) using
the five methods employed in GWW 
and even  (more clearly) with use of the
standard dispersion minimization method? Unfortunately not.
As seen from the bootstrap distribution on Fig.~\ref{figr05} the scatter
of the delays computed from trial runs is quite high and estimated
delays cover nearly the full range of the search. The mean of the bootstrap
delays is $-0.64$ years, and their standard deviation is $0.23$ years.
The large 
difference between the mean value of the bootstrap result and the
correct input value
($0.86-0.64 = 0.22$ years)
itself gives a strong
warning that the data set at hand is not large enough to 
estimate the time delay properly.

Details of the bootstrap
method as applied to the time delay estimation can be found in 
Pelt et al.~(\cite{Pelt96}). 
In GWW confidence limits for the time delay estimates
were computed using magnitude errors given by the observers. 
It is well known that the effective variability in data sets tends always 
to be larger due to
the possible outliers, unaccounted for systematic errors etc.
The bootstrap method tries to take all this into 
account and correspondingly
gives also larger, and we think more realistic, error bars
(see for instance Burud et al.~\cite{Burud00}). 
In the case of
HE 1104--1805 even the formal result $-0.86\pm 0.23$ years can not
be taken too seriously due to the bootstrap bias. 

\section{Discussion}
If we look carefully at all five time delay estimation statistics
used in GWW we see that, in principle, the results obtained 
can not very much differ from the results 
of the dispersion minimization method.
All statistics used can be looked upon as quadratic forms
with delay and weighting dependent coefficients. They all have a
free parameter (semiwidth of the bins or $\delta$) which controls smoothness
against bias of the statistics. Correspondingly, it is quite natural that
the $-0.85$ years feature occurs in all of them. 
The only important difference between 
the GWW methods and the dispersion minimization 
method is that
GWW uses a data ``cleaning'' which is based on an 
inappropriate idea of consistency. Hence it is not clear how 
trustworthy their analysis is.

We are far from claiming that the dispersion minimization method
is significantly better than any of those methods described in
GWW. Our basic emphasis is on the transparency of the method,
simplicity of its implementation and flexibility. As an explorative  
device it can effectively be used in most cases as a first trial. For a final
and definitive time delay estimation other methods which do not contain
any free parameters should be used. One such method 
(for relatively large data sets) is
the linear estimation method by Press et al.~\cite{Press92} with certain
extensions to take into account possible microlensing.
For small data sets probably
Bayesian methods (Loredo~\cite{Loredo92}) 
can be useful (with priors from
a large sample of variable quasars).

Now about the time delay for HE 1104-1805. If 
we want to consider the problem
as a full range search then it is quite obvious that 19 time points
with the particular spacing are absolutely insufficient to estimate the time
delay in this case. 
There is a wide range of delays which are not effectively 
covered by comparable time point pairs. Even if we look at 
another scenario where the acceptable time delay range 
is {\it a priori} restricted
to be from $-1.0$ years to $-0.65$ years, 
we can only say that there is a vague
hint in the data that the time delay can be somewhere in the range 
between $-0.9$ years and $-0.7$ years.
Unfortunately this precision is not enough for
useful physical applications.

From Fig.~\ref{figr03} we see that in order to obtain a realistic estimate
for the time delay for HE 1104-1805 we need new observations
with better sampling. Especially needed are observations which are
about half a year apart.   

\begin{acknowledgements}
We thank an anonymous referee for constructive criticism and numerous
suggestions. We thank also
Jan-Erik Ovaldsen and Jan Teuber for useful discussions. 
The work of J.P. was supported by the Estonian Science Foundation
(grant No. 4697) and by the Norwegian Research Foundation
(grant No. 142418/431).
  
\end{acknowledgements}

\end{document}